\documentclass[aps,prd,superscriptaddress,showpacs,preprint]{revtex4}
\usepackage{graphicx, bm}
\begin{document}
\draft
\title{Emissivity of neutrinos in supernova in a left-right\\
symmetric model}

\author{ A. Guti\'errez-Rodr\'{\i}guez}
\affiliation{\small Facultad de F\'{\i}sica, Universidad Aut\'onoma de Zacatecas\\
         Apartado Postal C-580, 98060 Zacatecas, M\'exico.}

\author{E. Torres-Lomas$^1$}

\author{A. Gonz\'alez-S\'anchez$^1$}

\date{\today}

\begin{abstract}

We calculate the emissivity due to neutrino-pair production in
$e^+e^-$ annihilation in the context of a left-right symmetric
model in a way that can be used in supernova calculations. We also
present some simple estimates which show that such process can act
as an efficient energy-loss mechanism in the shocked supernova
core. We find that the emissivity is dependent of the mixing angle
$\phi$ of the model in the allowed range for this parameter.
\end{abstract}


\vspace{5mm}

\maketitle


\section{Introduction}

The existence of neutrinos was postulated by Pauli in 1932 in
order to explain the observed continuous electron spectrum
accompanying nuclear beta decay. Based on the idea of Pauli, Fermi
\cite{Fermi,Fermi1} proposed the beta decay theory, while Bethe
and Peierls \cite{Bethe,Bethe1,Bethe2} predicted an extremely
small cross-section for the interaction of neutrino with matter,
and Gamow \cite{Gamow,Gamow1} and Pontecorvo \cite{Pontecorvo}
were the first to recognize the important role played by neutrinos
in the evolution of stars. The neutrino emission processes may
affect the properties of matter at high temperatures, and hence
affect stellar evolution.

Neutrino emission is known to play an important role in stellar
evolution, especially in the late stages when the rate of
evolution is almost fully dependent on energy loss via neutrinos.
This refers to the stage of steady burning prior to the implosion
of the stellar core, to the process of catastrophic core-collapse,
and to the cooling of the neutron star which is formed.

The explosion energy of the core-collapse is typically
$10^{53}\hspace{1mm} erg$ which makes it one of the most
impressive violent events in the universe. This energy comes from
the explosion of the progenitor star, and only partly manifests
itself in the shock wave that is launched somewhere at the
boundary between the iron core of mass $M_{Fe}= (1.2-2)M_{\odot}$
and the inner most regions, collapsing into a neutron star. Even
when the mechanism of the core-collapse is not yet understood in
great detail, the most distinctive feature is the enormous energy
of $ (3-5)\times 10^{53}\hspace{1mm} erg =
(10-15)\%\hspace{1mm}M_{Fe}c^{2}$ radiated in the form of
neutrinos and antineutrinos of all flavors $(\nu_e, \nu_\mu,
\nu_\tau)$ during a burst of about 10 seconds. While such
neutrinos were first observed in the supernova SN1987A, various
observatories running or under design, like next generation
large-size detectors, could provide us with the luminosity curve
from a future (extra)galactic explosion and/or the observation of
relic neutrinos from past supernovae. To disentangle the
information from such neutrino signals represents a challenging
task, since the crucial information from the explosion phenomenon
and neutrino properties such as the neutrino hierarchy and the
third neutrino mixing angle are intertwined.

The detection of neutrinos from SN1987A by the Kamiokande II
\cite{Hirata} and Irvine-Michigan-Brookhaven \cite{Bionta}
detectors confirmed the standard model of core-collapse (type II)
supernovae \cite{Arnett,Burrows} and provided a laboratory to
study the properties of neutrinos
\cite{Schramm,Raffelt,Gaemers,Grifols,Gandhi,Turner} and exotic
particles such as axions \cite{Turner1}. The collapse of stellar
iron-core into a neutron star is preceded by a high-power pulse of
neutrino emission. In general, a bolometric neutrino light curve
that includes all the neutrino and antineutrino flavors consists
of two parts. In the first part ($t < 5$ $s$) the non-thermal
neutrino emission is dominated by the electron neutrinos $\nu_{e}$
produced by the non-thermal neutronization. For $t \approx  0.5$
$s$, the core is transparent to $\nu_{e}$ emitted due to electrons
captures by nuclei and free protons. By this time, the mean
individual $\nu_{e}$ energy becomes $10-20$ $MeV$. Thus, the
non-thermal neutrinos carry away approximately a small fraction of
the total available energy $E_{\nu tot} = (3-5)\times 10^{53}$
erg. Nearly $90\%$ of this energy is emitted in the regime of
thermal emission once the innermost region of the core becomes
opaque to all the neutrino flavors, which get decoupled from the
stellar mass at a surface so-called neutrinosphere.

Therefore, one of the crucial parameters which strongly affect the
stellar evolution is the cooling rate. During their lifetime,
stars can emit energy in the form of electromagnetic or
gravitational waves, and a flux of neutrinos. However, in late
stages a star mainly looses energy through neutrinos, and this is
quite independent of the star mass. In fact, white dwarfs and
supernovae, which are the evolution end points of stars formed
from very different masses, have cooling rates largely dominated
by neutrino production. An accurate determination of neutrino
emission rates is therefore mandatory in order to perform a
careful study of the final branches of star evolutionary tracks.
In particular, a change in the cooling rates at the very last
stage of massive star evolution could perceptibly affect the
evolutionary time scale and the iron core configuration at the
onset of the supernova explosion whose triggering mechanism still
waits a full theoretical understanding \cite{Janka}.

The energy loss rate due to neutrino emission receives
contributions from both weak nuclear reactions and purely leptonic
processes. However, for the large values of density and
temperature which characterize the final stage of stellar
evolution, the latter are largely dominant, and are mainly
produced by four possible interaction mechanisms: $e^+ e^- \to \nu
\bar \nu$ (pair annihilation), $\gamma e^\pm \to  e^\pm \nu \bar
\nu$ ($\nu$-photoproduction), $\gamma^* \to  \nu \bar \nu$
(plasmon decay), $e^\pm Z \to e^\pm Z \nu \bar \nu$
(bremsstrahlung on nuclei). These mechanisms play an important
role in astrophysics and cosmology and have been considered by
many authors in various theories of weak interactions.

Actually  these processes are the dominant cause of the energy
loss rate in different regions in a density-temperature plane. For
very large core temperature, $T\gtrsim 10^{9}$ $^oK$, and not
excessively high values of density, pair annihilations are most
efficient, while $\nu$ photoproduction gives the leading
contribution for $10^{8}$ $^oK$ $ \lesssim T\lesssim 10^{9}$ $^oK$
and relatively low density, $\rho \lesssim 10^5$ $g$ $cm^{-3}$.
These are the typical ranges for very massive stars in their late
evolution.

Our main objetive in this paper is to provide suitable expressions
for the emissivity of pair production of neutrinos via the process
$e^+ e^- \to \nu \bar \nu$ in the context of a Left-Right
Symmetric Model (LRSM)
\cite{Pati,R.N.Mohapatra,R.N.Mohapatra1,G.Senjanovic,G.Senjanovic1,M.Maya,R.Huerta,Polak}
and in a form which can be easily incorporated into realistic
supernova models to evaluate the energy lost in the form of
neutrinos.

The amplitude of transition ${\cal M}$ in the context of the LRSM
can be written as a function of the mixing angle $\phi$ between
$W^3_L$, $W^3_R$ and $B$ bosons of the model to give the physical
$Z_1$ and $Z_2$ and the photon, being $\phi$ the only extra
parameter besides the standard model parameters. For which in this
paper we choice the Left-Right symmetric model
\cite{M.Maya,R.Huerta,Polak} to calculate the emissivity of
neutrinos in supernova.

This paper is organized as follows: In Sect. II we present the
calculation of the transition amplitude of the process $e^+ e^-
\to \nu \bar \nu$ in the context of a left-right symmetric model.
In Sect. III we calculate the emissivity and, finally, we give our
results and conclusions in Sec. IV.

\section{Cross Section of the Process $e^+ +e^- \rightarrow \nu +\bar{\nu}$}

In this section we obtain the cross section for the $Z$ exchange
process

\begin{equation}
e^+(p_1)+e^-(p_2)\rightarrow \bar \nu (k_1, \lambda_1)+ {\nu}(k_2,
\lambda_2),
\end{equation}

\noindent i.e., in the limit of a four-fermion electroweak
interaction no electromagnetic radiative corrections. Here the
$k_i$ and $p_i$ are the particle momenta and $\lambda$ is the
helicity of the neutrino. We recall that within the context of the
standard theory, a neutrino interaction eigenstate ($\nu_L$ or
$\nu_R$) is a superposition of helicity ($\lambda$) eigenstates
$\nu_{\pm}$, where $\lambda=\bf \sigma\cdot p=\pm $1. For a
relativistic particle, this translates into the statement that a
$\nu_L$ is predominantly in the $\lambda=-1$ state and a $\nu_R$
is predominantly in the $\lambda=+1$ state, with small admixtures
of the opposite helicity of the order $m/E_\nu$.

The amplitude of transition for the process (1) is given by

\begin{equation}
{\cal M}=\frac{g_Z^2}{2M_Z^2}\left[\bar{u}\left(k_2,
\lambda_2\right)\gamma^\mu\frac{1}{2}\left(a g_V^\nu-b g_A^\nu
\gamma_5\right)v\left(k_1,
\lambda_1\right)\right]\left[\bar{v}\left(p_1\right)\gamma_\mu\frac{1}{2}\left(a
g_V^e-b g_A^e \gamma_5\right)u\left(p_2\right)\right],
\end{equation}

\noindent where the constant $a$ and $b$ depend only on the
parameters of the LRSM model \cite{M.Maya,R.Huerta,Polak}

\begin{equation} a=\cos{\phi}-\frac{\sin{\phi}}{\sqrt{\cos2\theta_W}} \hspace{5mm} \mbox{and}
\hspace{5mm} b=\cos{\phi}+\sqrt{\cos2\theta_W}\sin{\phi},
\end{equation}

\noindent where $\phi$ is the mixing parameter of the LRSM
\cite{M.Maya,R.Huerta,Polak}, $u$ and $v$ are the usual Dirac
spinors, and the electron and positron helicity indexes have been
suppressed since they will be averaged over. We then write

\begin{equation}
\frac{1}{2}\times\frac{1}{2}|{\cal
M}|^2=\frac{G_F^2}{8}N^{\mu\nu}E_{\mu\nu},
\end{equation}

\noindent where

\begin{eqnarray}
N^{\mu\nu}&=&\frac{1}{4}Tr[( k\llap{/}_{2}+m_\nu)(1+\gamma^5
s\llap{/}_{2})\gamma^\mu(a-b\gamma_5) ( k\llap{/}_{1}
-m_\nu)(1+\gamma^5 s\llap{/}_{1})\gamma^\nu(a-b\gamma_5)],\\
E_{\mu\nu}&=&\frac{1}{4}Tr[( p\llap{/}_{2}
+m_e)\gamma_\mu(ag^e_V-bg^e_A\gamma_5) ( p\llap{/}_{1}
-m_e)\gamma_\nu(ag^e_V-bg^e_A\gamma_5)].
\end{eqnarray}

Here $s_1$ and $s_2$ are the spin four-vectors associated with the
antineutrino and neutrino respectively, while  $m_\nu$ and $m_e$
are the neutrino and electron mass. These spin vectors satisfy the
Lorentz invariant conditions

\begin{equation}
s_i\cdot s_i=-1; \hspace{3mm} s_i\cdot k_i=0;
\end{equation}

\noindent and for a relativistic neutrino, the additional
constraint

\begin{equation}
{\bf s}_i \parallel \lambda_i {\bf k}_i \hspace{3mm} \mbox {for}
\hspace{2mm} i=1,2
\end{equation}

\noindent holds, where

\begin{equation}
k^\mu=(E_\nu, k{\hat{\bf k}}),
\end{equation}

\noindent with $\hat{\bf k}$ being a unit vector along the
three-momentum of the neutrino.

We now introduce two four-vectors associated with the neutrino
pair

\begin{equation}
K^\mu_1= k^\mu_1+m_\nu s^\mu_1; \hspace{3mm} K^\mu_2= k^\mu_2-
m_\nu s^\mu_2.
\end{equation}

In conjunction with the properties given in Eqs. (7) and (8),
these will allow us to write the amplitude squared for the process
under consideration in a compact and physically revealing form. As
a first step towards this, we note that the spin vector may be
expressed as

\begin{equation}
s^\mu=\frac{\lambda}{m_\nu}(k, E_\nu{\hat{\bf k}}).
\end{equation}

Using this and Eq. (10), we write

\begin{equation}
K_1=\eta_1(1, {\hat{\bf k}}); \hspace{3mm} K_2=\eta_2(1,
-{\hat{\bf k}});
\end{equation}

\noindent with

\begin{equation}
\eta_1=E_\nu + (E^2_\nu -m^2_\nu)^{1/2}; \hspace{3mm} \eta_2=E_\nu
- (E^2_\nu -m^2_\nu)^{1/2}.
\end{equation}

Note that for $m_\nu << E_\nu$ we have

\begin{equation}
\eta_1 \approx 2E_\nu; \hspace{3mm} \eta_2 \approx
\frac{m^2_\nu}{2E_\nu}.
\end{equation}

We now evaluate the traces given in Eqs. (5) and (6) and the
contraction $N^{\mu\nu} E_{\mu\nu}$ is given by

\begin{eqnarray}
N^{\mu\nu}E_{\mu\nu}=&8&(a^2+b^2)
\left\lbrace\left[ a^2(g^e_V)^2+b^2(g^e_A)+4\frac{a^2b^2}{(a^2+b^2)}g^e_Vg^e_A\right](p_1\cdot K_1)(p_2\cdot K_2)  \right.\nonumber\\
&+&\left[ \phantom{\frac{a}{b}}a^2(g^e_V)^2+b^2(g^e_A)-4\frac{a^2b^2}{(a^2+b^2)}g^e_Vg^e_A\right] (p_1\cdot K_2) (p_2\cdot K_1)\nonumber\\
&+&\left.\left[a^2(g^e_V)^2-b^2(g^e_A)\phantom{\frac{a}{b}}\right]m_e^2(K_1\cdot K_2)\right\rbrace, \nonumber\\
\end{eqnarray}

\noindent where $g^e_V=-\frac{1}{2}+2\sin^2\theta_W$ and
$g^e_A=-\frac{1}{2}$. From this expression and Eqs. (12) and (14)
above, we see that the amplitude of transition vanishes for
massless neutrino, as expected. Furthermore, Eq. (15) is taken as
the usual weak pair production amplitude with the replacement $K_i
\to k_i$.

From Eqs. (4) and (15) the explicit form for the squared
transition amplitude is

\begin{eqnarray}
\frac{1}{2}\times\frac{1}{2}|{\cal M}|^2=&G_F^2&
(a^2+b^2)\left\lbrace\left[ a^2(g^e_V)^2+b^2(g^e_A)+4\frac{a^2b^2}{(a^2+b^2)}g^e_Vg^e_A\right] (p_1\cdot k_1) (p_2\cdot k_2)\right.\nonumber\\
&+&\left[ \phantom{\frac{a}{b}}a^2(g^e_V)^2+b^2(g^e_A)-4\frac{a^2b^2}{(a^2+b^2)}g^e_Vg^e_A\right] (p_1\cdot k_2)(p_2\cdot k_1)\nonumber\\
&+&\left.\left[a^2(g^e_V)^2-b^2(g^e_A)\phantom{\frac{a}{b}}\right]m_e^2(k_1\cdot
k_2)\right\rbrace,
\end{eqnarray}

\noindent where the contribution of the parameters of the LRSM is
contained in the constants $a$ and $b$. Upon evaluating the limit
when the mixing angle $\phi = 0$ and $a=b=1$, Eq. (16) is thus
reduced to the expression to the amplitude given in Refs.
\cite{Yakovlev,Esposito1,Esposito2,Armando,Misiaszek,Dicus,Ellis}.

\section{Calculation of Emissivity}

In this section, we calculate the emissivity associated with
neutrino pair production by using Eq. (16). The formula of the
emissivity is given by
\cite{Yakovlev,Esposito1,Esposito2,Misiaszek,Lenard}

\begin{equation}
Q_{\nu\bar\nu}=\frac{4}{(2\pi)^8}\int\frac{d^3\mathbf{p}_1}{2E_1}\frac{d^3\mathbf{p}_2}{2E_2}
\frac{d^3\mathbf{k}_1}{2\epsilon_1}\frac{d^3\mathbf{k}_2}{2\epsilon_2}(E_1+E_2)F_1F_2
\delta^{(4)}(p_1+p_2-k_1-k_2)|{\cal M}|^2,
\end{equation}

\noindent where the quantities $F_{1,2}=[1+\exp(E_{e^-}\pm
\mu_{e^-} )/T]^{-1}$ are the Fermi-Dirac distribution functions
for $e^{\pm}$, $\mu_e$ is the chemical potential for the electrons
and $T$ is the temperature (we take $K_B=1$ for the Boltzmann
constant).

From the transition amplitude Eq. (16) and the formula of the
emissivity Eq. (17) we obtain

\begin{equation}
Q^{[1]}_{\nu\bar\nu}=G_F^2(a^2+b^2)\left[a^2\left(
g^e_V\right)^2+b^2 \left( g^e_A\right)^2+\frac{4a^2b^2}{a^2+b^2}
g^e_V g^e_A\right]I_1,
\end{equation}

\noindent where $I_1$ is explicitly given by

\begin{equation}
I_1=\frac{4}{(2\pi)^8}\int\frac{d^3\mathbf{p}_1}{2E_1}\frac{d^3\mathbf{p}_2}{2E_2}
\frac{d^3\mathbf{k}_1}{2\epsilon_1}\frac{d^3\mathbf{k}_2}{2\epsilon_2}(E_1+E_2)F_1F_2
\delta^{(4)}(p_1+p_2-k_1-k_2)(p_1\cdot k_1)(p_2\cdot k_2).
\end{equation}

The intregration can be performed by using the Lenard formula,
namely \cite{Lenard}

\begin{eqnarray}
\int\frac{d^3\mathbf{k}_1}{2\epsilon_1}\frac{d^3\mathbf{k}_2}{2\epsilon_2}k_1^\alpha k_2^\beta\delta^{(4)}(p_1+p_2-k_1-k_2)
&=&\frac{\pi}{24}\left[g^{\alpha\beta}(p_1+p_2)^2+2(p_1^\alpha+p_2^\alpha)(p_1^\beta+p_2^\beta)\right]\nonumber\\
&&\cdot \Theta\left[(p_1+p_2)^2\right],
\end{eqnarray}

\noindent thus Eq. (19) takes the form

\begin{equation}
I_1=\frac{1}{24(2\pi)^7}\int\frac{d^3\mathbf{p}_1}{E_1}\frac{d^3\mathbf{p}_2}{E_2}(E_1+E_2)F_1F_2\left[
3m_e^2(p_1\cdot p_2)+2(p_1\cdot p_2)^2+m_e^4 \right ].
\end{equation}

In a similar way for the second and third term of Eq. (16), we
obtain

\begin{eqnarray}
Q^{[2]}_{\nu\bar\nu}&=&G_F^2(a^2+b^2)\left[a^2\left(
g^e_V\right)^2+b^2 \left( g^e_A\right)^2-\frac{4a^2b^2}{a^2+b^2}
g^e_V g^e_A\right]I_2,\\
Q^{[3]}_{\nu\bar\nu}&=&G_F^2(a^2+b^2)\left[a^2\left(
g^e_V\right)^2-b^2 \left( g^e_A\right)^2\right]m_e^2 I_3,
\end{eqnarray}

\noindent where

\begin{eqnarray}
I_2&=&I_1=\frac{1}{24(2\pi)^7}\int\frac{d^3\mathbf{p}_1}{E_1}\frac{d^3\mathbf{p}_2}{E_2}(E_1+E_2)F_1F_2\left[
3m_e^2(p_1\cdot p_2)+2(p_1\cdot p_2)^2+m_e^4 \right ],\\
I_3&=&\frac{1}{(2\pi)^7}\int\frac{d^3\mathbf{p}_1}{2E_1}\frac{d^3\mathbf{p}_2}{2E_2}
(E_1+E_2)F_1F_2 \left[ m_e^2+(p_1\cdot p_2) \right].
\end{eqnarray}

The calculation of the emissivity can be more easily performed by
expressing the latest integrals in terms of the Fermi integral,
which is defined  as \cite{Misiaszek}

\begin{equation}
G_s^{\pm}=\frac{1}{\alpha^{3+2s}}\int_\alpha^\infty
x^{2s+1}\frac{\sqrt{x^2-\alpha^2}}{1+e^{x\pm\beta}}dx,
\end{equation}

\noindent where $\alpha=\frac{m_e}{KT}$, $\beta=\frac{\mu_e}{KT}$
and $x=\frac{E}{KT}$.

With these definitions, Eq. (26) becomes

\begin{equation}
G_s^{\pm}=\frac{1}{m_e^{3+2s}}\int_{m_e/KT}^\infty
E^{2s+1}\frac{\sqrt{E^2- m_e^2}}{1+e^{(E\pm\mu_e)/KT}}dE,
\end{equation}

\noindent therefore

\begin{equation}
\int_{m_e/KT}^\infty E^{n}\frac{\sqrt{E^2-
m_e^2}}{1+e^{(E\pm\mu_e)/KT}}dE=m_e^{n+2}G_{\frac{n-1}{2}}^{\pm},
\end{equation}

\begin{equation}
\int_{m_e/KT}^\infty E^{n+1}\frac{\sqrt{E^2-
m_e^2}}{1+e^{(E\pm\mu_e)/KT}}dE=m_e^{n+3}G_{\frac{n}{2}}^{\pm},
\end{equation}

\begin{equation}
\int_{m_e/KT}^\infty E^{n+2}\frac{\sqrt{E^2-
m_e^2}}{1+e^{(E\pm\mu_e)/KT}}dE=m_e^{n+4}G_{\frac{n+1}{2}}^{\pm}.
\end{equation}

From (28), (29) and (30), Eqs. (24) and (25) are expressed as

\begin{eqnarray}
I_1^{nm}&=&\frac{m_e^{n+m+8}}{6(2\pi)^5}\left[3G_{\frac{n}{2}}^{-}G_{\frac{m}{2}}^{+}+2G_{\frac{n+1}{2}}^{-}G_{\frac{m+1}{2}}^{+}+G_{\frac{n-1}{2}}^{-}G_{\frac{m-1}{2}}^{+}\right. \nonumber\\
&&\left.
+\frac{4}{9}\left(G_{\frac{n+1}{2}}^{-}-G_{\frac{n-1}{2}}^{-}\right)\left(G_{\frac{m+1}{2}}^{+}-G_{\frac{m-1}{2}}^{+}\right)\right],\\
I_3^{nm}&=&\frac{m_e^{n+m+6}}{(2\pi)^5}\left[G_{\frac{n-1}{2}}^{-}G_{\frac{m-1}{2}}^{+}+G_{\frac{n}{2}}^{-}G_{\frac{m}{2}}^{+}\right].
\end{eqnarray}

Therefore, Eqs. (18), (22) and (23) are explicitly

\begin{eqnarray}
Q^{[1]}_{\nu\bar\nu}&=&G_F^2(a^2+b^2)\left[a^2\left(
g^e_V\right)^2+b^2 \left( g^e_A\right)^2+\frac{4a^2b^2}{a^2+b^2}
g^e_V g^e_A\right]\left[ I_1^{10}+I_1^{01}\right],\\
Q^{[2]}_{\nu\bar\nu}&=&G_F^2(a^2+b^2)\left[a^2\left(
g^e_V\right)^2+b^2 \left( g^e_A\right)^2-\frac{4a^2b^2}{a^2+b^2}
g^e_V g^e_A\right]\left[ I_2^{10}+I_2^{01}\right],\\
Q^{[3]}_{\nu\bar\nu}&=&G_F^2(a^2+b^2)\left[a^2\left(
g^e_V\right)^2-b^2 \left( g^e_A\right)^2\right]m_e^2\left[
I_3^{10}+I_3^{01}\right].
\end{eqnarray}

Finally, the expression for the emissivity of neutrino pair
production via the process $e^+ e^- \to \nu \bar \nu$ in the
context of a left-right symmetric model is given by

\begin{equation}
Q^{LRSM}_{\nu\bar\nu}\left(\phi,\beta\right)=Q_{\nu\bar\nu}^{[1]}\left(\phi,\beta\right)+Q_{\nu\bar\nu}^{[2]}\left(\phi,\beta\right)+Q_{\nu\bar\nu}^{[3]}\left(\phi,\beta\right),
\end{equation}

\noindent where the dependence of the $\phi$ mixing parameter of
the LRSM is contained in the constants $a$ and $b$, while the
dependence of the $\beta$ degeneration parameter is contained in
the Fermi integrals $G_s^{\pm}$.

\section{Results and Conclusions}

The numerical result on the emissivity of neutrino pair production
via the process $e^+ e^- \to \nu \bar \nu$ as a function of the
mixing angle $\phi$ and the generation parameter $\beta$ are
present in this section. For our analysis we consider the
following data: the Fermi constant $G_F=1.166\times10^{-5}$
$GeV^{-2}$, angle of Weinberg $\sin^2\theta_W=0.223$ and the
electron mass $m_e=0.51$ $MeV$, thereby obtaining the emissivity
of the neutrinos
$Q^{LRSM}_{\nu\bar\nu}=Q^{LRSM}_{\nu\bar\nu}\left(\phi,\beta\right)$.

For the mixing angle $\phi$ of the left-right symmetric model, we
use the reported data of A. Guti\'errez-Rodr{\'i}guez, {\it et
al.} \cite{A.Gutierrez2}:

\begin{equation}
-1.6\times 10^{-3}\leq\phi \leq 1.1\times 10^{-3},
\end{equation}

\noindent with a $90\%$ C.L. Other limits on the mixing angle
$\phi$ reported in the literature are given in Refs.
\cite{A.Gutierrez3,M.Maya,Polak,Adriani}.

In Figure 1 we show the emissivity as a function of the
degeneration parameter $\beta$ for several values representative
of the mixing angle $\phi=-0.0016, 0, 0.0011$. We observe that
emissivity decreases when $\beta$ increases, which is due to the
reduction in the number of positrons available necessary to cause
the collision. Also we see that the emissivity is affected by the
$\phi$ parameter. There are other effects which may change the
emissivity, for example, the radiative corrections at one-loop
level.

To analyze the effects of the $\phi$ parameter of the left-right
symmetric model on the emissivity
$Q^{LRSM}_{\nu\bar\nu}\left(\phi, \beta\right)$ of the neutrinos,
in Fig. 2 we show the ratio
$\frac{Q^{LRSM}_{\nu\bar\nu}\left(\phi,
\beta\right)}{Q^{SM}_{\nu\bar\nu}\left(\beta\right)}$ (ratio of
emissivity calculated in the LRSM to the emissivity calculated in
the SM \cite{Dicus}), as a function of the $\phi$ parameter for
several values representative of degeneration parameter $\beta=0,
5, 12$.

According to collapse theories, the full energy loss in a stellar
collapse is $L_{\nu \bar \nu}=Q_{\nu \bar \nu} \cdot V_{star}
\cdot t_{col}\approx 10^{53}\hspace{1mm}erg$, where $V_{star}$ is
the volume of a star and $t_{col}$ the collapse time. Therefore,
if we assume that $t_{col}\sim 1-10\hspace{1mm}s$, the stellar
collapse temperature $T_{col}\sim (1.1-1.4) \times 10^{11}
\hspace{1mm}^oK$ will be obtained. On the contrary, if we know the
exact stellar collapse temperature, the collapse time can be
obtained. For instance, the value $T_{col}\sim \times 10^{12}
\hspace{1mm}^oK$ corresponds to $t_{col}\sim 10^{-7}
\hspace{1mm}s$.

In summary, we have analyzed the effects of the mixing angle
$\phi$ of a left-right symmetric model on the emissivity of the
neutrinos via the process $e^+ e^- \to \nu \bar \nu$. We find that
the emissivity is dependent of the mixing angle $\phi$ of the
model in the allowed range for this parameter. As expected, in the
limit of vanishing $\phi$ we recover the expression for the
emissivity $Q^{SM}_{\nu\bar\nu}\left(\beta\right)$ for the SM
previously obtained in the literature
\cite{Armando,Yakovlev,Esposito1,Esposito2,Misiaszek}. In
addition, the analytical and numerical results for the emissivity
have never been reported in the literature before and could be of
relevance for the scientific community.

\vspace{1cm}

\begin{center}
{\bf Acknowledgments}
\end{center}

This work was supported by CONACyT, SNI and PROMEP (M\'exico).


\newpage


\begin{figure}[t]
\centerline{\scalebox{0.8}{\includegraphics{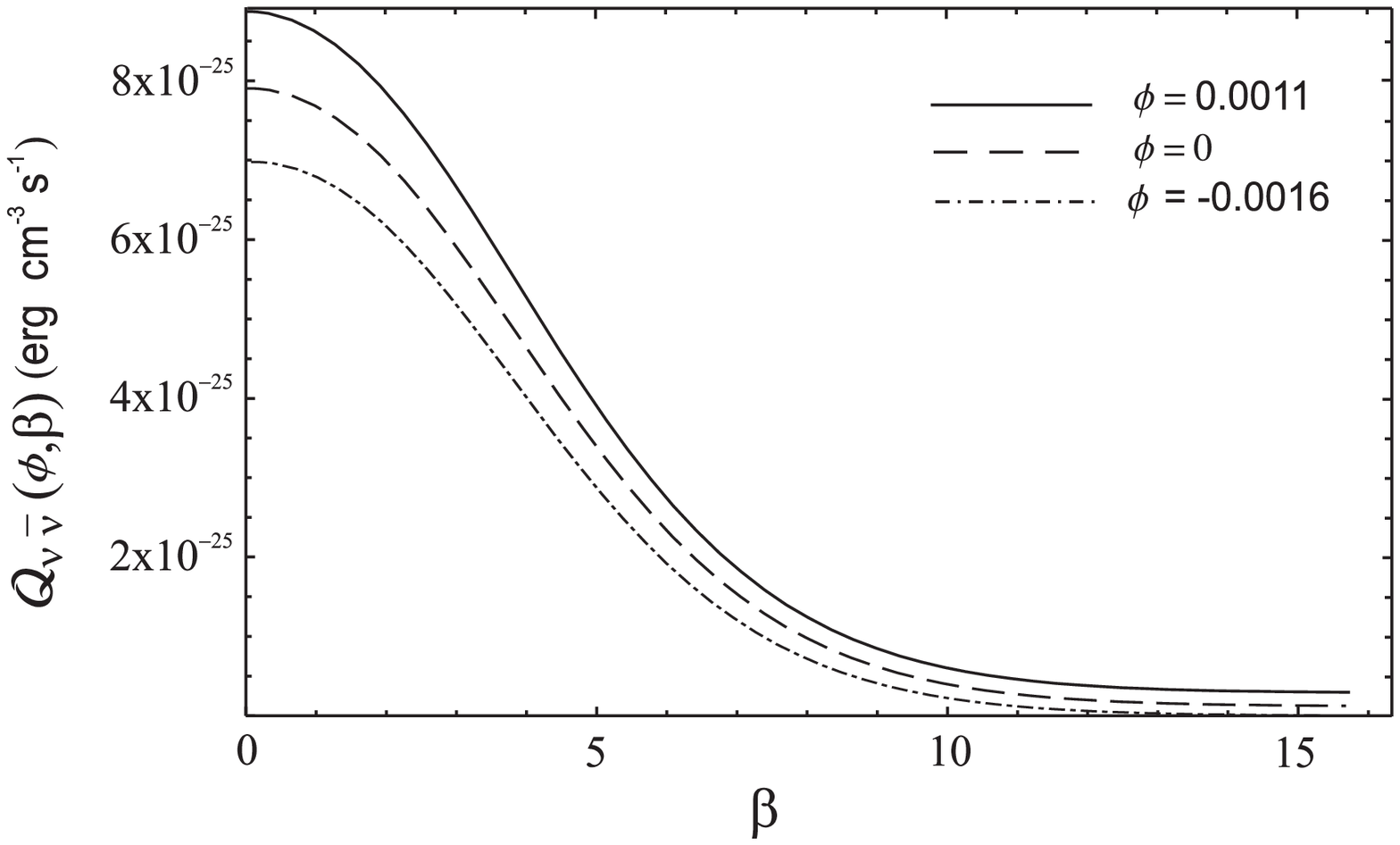}}}
\caption{ \label{fig:gamma} The emissivity for $e^+ e^- \to \nu
\bar \nu$ as a function of degeneration parameter $\beta$ for
$\phi=-0.0016, 0, 0.0011$.}
\end{figure}

\begin{figure}[t]
\centerline{\scalebox{1.3}{\includegraphics{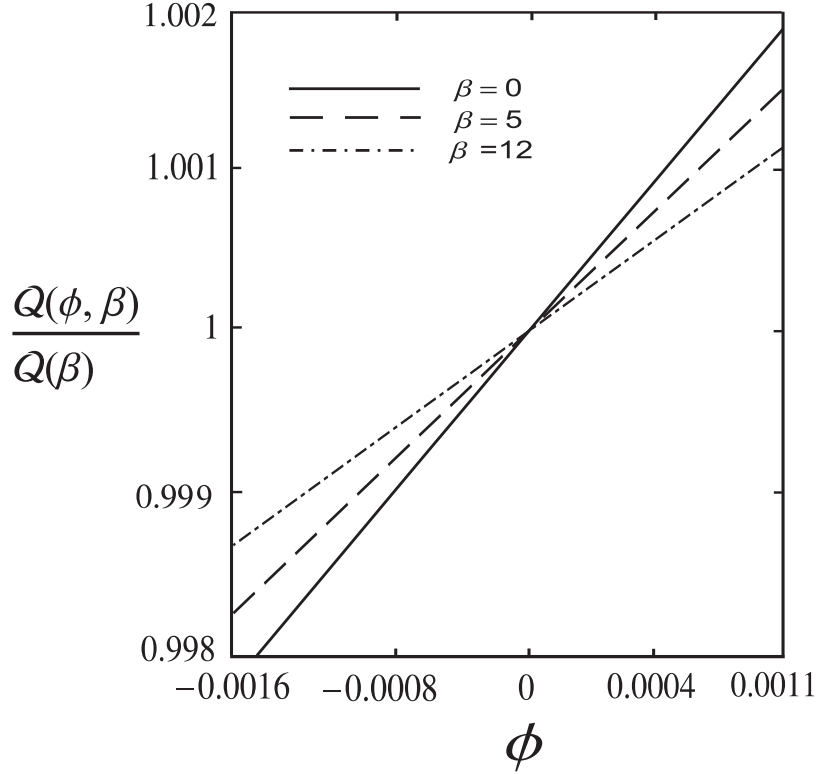}}}
\caption{ \label{fig:gamma} Plot of ratio
$\frac{Q_{\nu\bar\nu}\left(\phi,
\beta\right)}{Q_{\nu\bar\nu}\left(\beta\right)}$, as a function of
mixing angle $\phi$ for $\beta = 0, 5, 12$.}
\end{figure}

\end{document}